\documentclass{PoS}

\title{The Cherenkov Telescope Array}

\ShortTitle{The Cherenkov Telescope Array}

\author{\speaker{Daniel Mazin} for the CTA Consortium\footnote{for the CTA Consortium list see PoS(ICRC2019)1177} \\
        ICRR, University of Tokyo, Kashiwa-no-ha 5-1-5, Kashiwa-shi, 277-8582 Chiba, Japan \& \\
        MPI for physics, F\"ohringer Ring 6, 80805 Munich, Germany\\
        E-mail: \email{mazin@icrr.u-tokyo.ac.jp} 
}

\abstract{
The Cherenkov Telescope Array (CTA) is the next generation ground-based
observatory for gamma-ray astronomy at very-high energies. It will be capable
of detecting gamma rays in the energy range from 20 GeV to more than 300 TeV
with unprecedented precision in energy and directional reconstruction. With
more than 100 telescopes of three different types it will be located in the northern hemisphere at La
Palma, Spain, and in the southern at Paranal, Chile. 
CTA will be one of the largest astronomical infrastructures in the world with open data access
and it will address questions in astronomy, astrophysics and fundamental physics
in the next decades. In this presentation we will focus on the status of the CTA construction,
the status of the telescope prototypes and highlight some of the physics perspectives. }

\FullConference{36th International Cosmic Ray Conference -ICRC2019-\\
		July 24th - August 1st, 2019\\
		Madison, WI, U.S.A.}

\begin{document}

\section{Introduction}
Ground-based gamma-ray astronomy --- imaging the universe at very high energies
(VHE) above tens of GeV and covering several decades of the electromagnetic
spectrum --- is a young branch of astronomy that has developed very rapidly
since the detection of the first cosmic VHE source in 1989 \cite{weekesCrab}. 
The Cherenkov Telescope Array (CTA) Consortium formed in 2008 to develop a
concept for the first major open observatory for this waveband, motivated by
the success of existing imaging atmospheric Cherenkov telescopes (IACTs) such
as H.E.S.S., MAGIC and VERITAS. These instruments have demonstrated that
observations at these extreme energies are not only technically viable and
competitive in terms of precision and sensitivity, but also highly rewarding
and with broad scientific impact. This success has resulted in a rapid growth
of the interested scientific community.
An overview of the Cherenkov technique with IACTs can be found in, e.g., \cite{deNauroisMazin:2015}
and the status of the field is discussed in, e.g., \cite{Degrange:2015}.

\section{CTA telescopes and layouts}

%Here I will describe the telescope types and layouts
\begin{figure}
\centering
\includegraphics[width=1.0\textwidth]{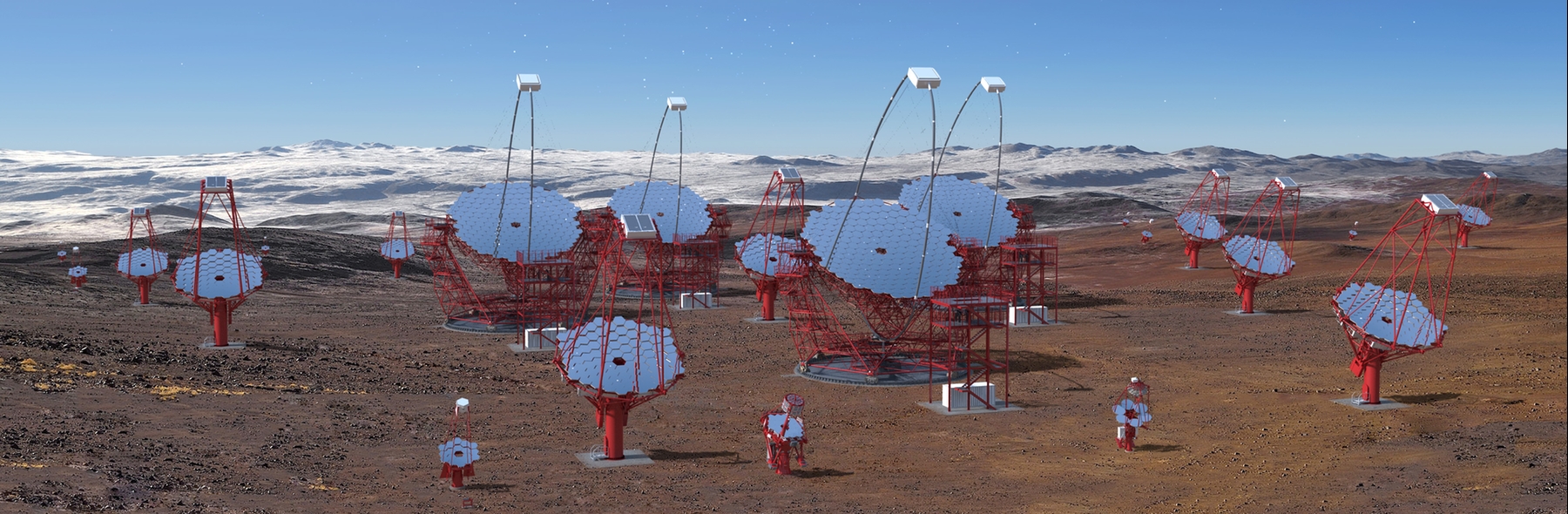}     % includes figure foo.eps
\includegraphics[width=1.0\textwidth]{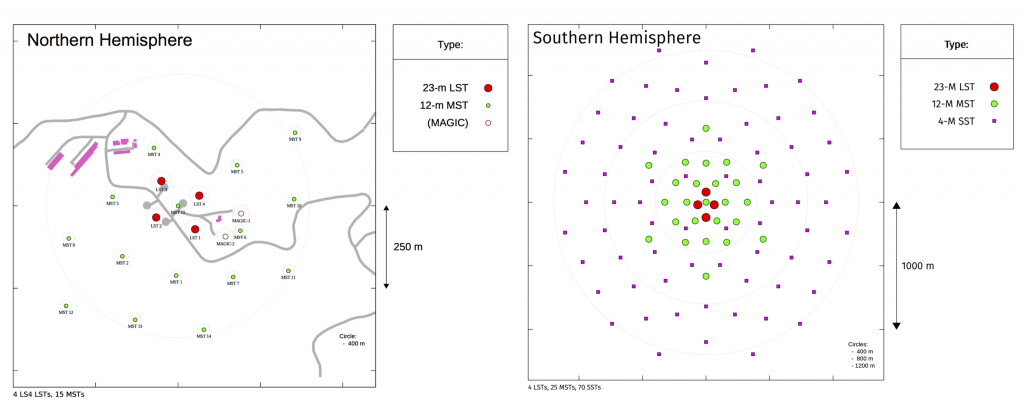}     % includes figure foo.eps
\caption{Artists view on the Southern CTA array (top) and array layout in the Northern (left)
and Southern (right) hemispheres (bottom), respectively. }
\label{fig:cta}
\end{figure}

CTA (see Figure~\ref{fig:cta}, top) will take the VHE gamma
astronomy to a next level, by constructing about 100 telescopes (compared to a
dozen of currently existing ones) and operating them as an open observatory.
CTA will be built on two sites, one in the northern and one in the southern
hemisphere to provide full sky coverage.
Rather than deploying one type of Cherenkov telescope on a regular grid, the
CTA arrays (Figure~\ref{fig:cta}, bottom) use a graded approach:
\begin{itemize}
\item The lowest energies are covered by an arrangement of 4 Large-Sized telescopes (LSTs), capable of detecting gamma rays down to 20 GeV.
\item The 0.1 to 10 TeV range is covered by a larger array of 25 (south) or 15 (north) Medium-Sized Telescopes (MSTs).
\item The highest energy gamma rays are detected by a multi-km$^2$ array of 70 Small-Sized Telescopes (SSTs), in the south.
\end{itemize}

\begin{figure}
\centering
\includegraphics[width=1.0\textwidth]{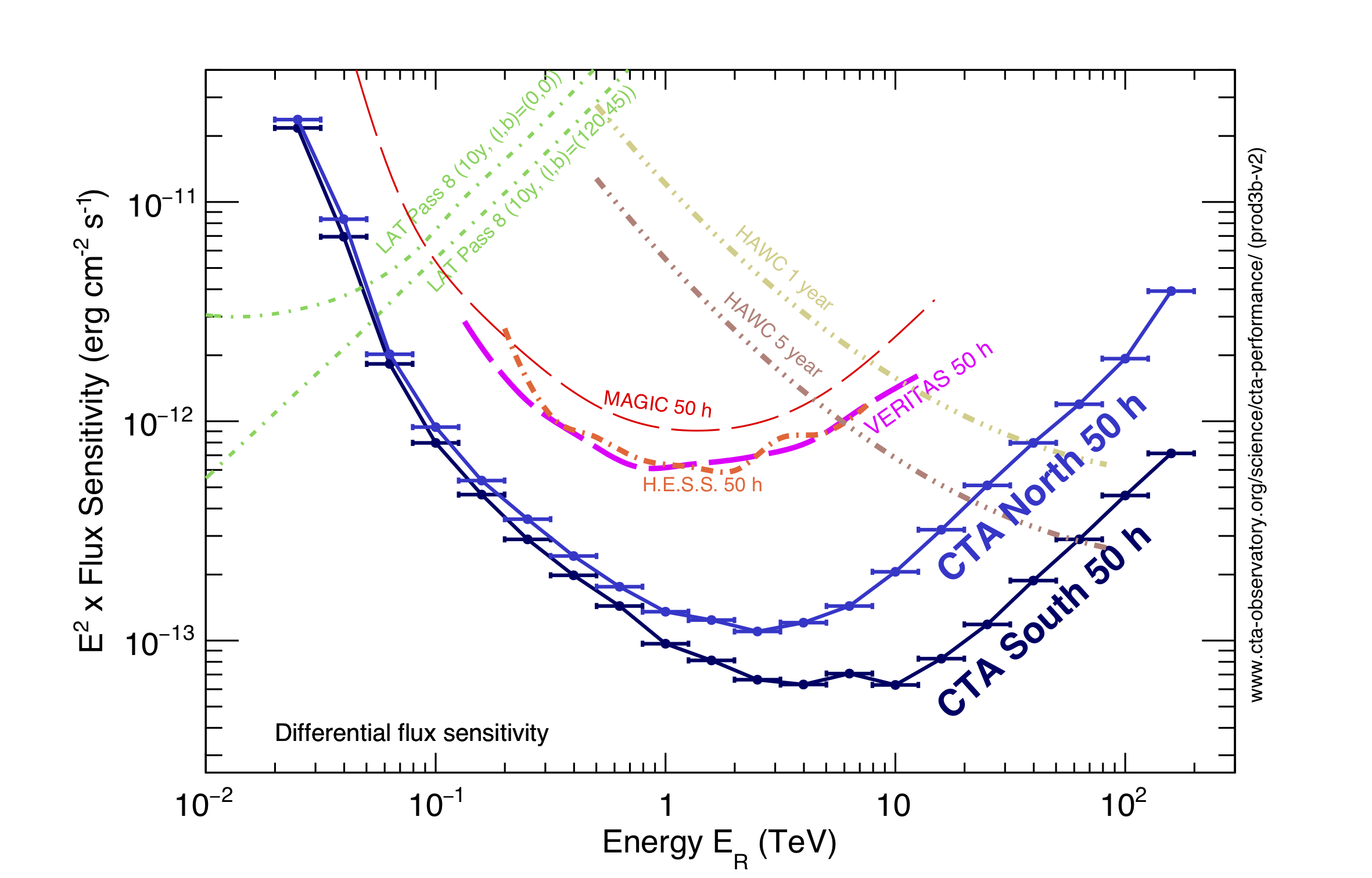}     % includes figure foo.eps
\caption{CTA differential flux sensitivity. 
Please find more performance characteristics at https://www.cta-observatory.org/science/cta-performance/.}
\label{fig:sensi}
\end{figure}

The SSTs telescopes are only
foreseen for the Southern array, since the highest energies are most relevant
for the study of Galactic sources, of which a larger fraction is observable from the Southern Hemisphere. 
The use of three different sizes of
telescopes proved to be the most cost-effective solution, and it allows each
telescope type to be optimised for a specific energy range. 
A preliminary flux sensitivity for 50\,h of observations with the two arrays is shown in Figure~\ref{fig:sensi}, 
based on Monte Carlo simulations. Note the improved energy coverage and sensitivity compared to the existing IACTs.
An interested reader is advised see more details on the CTA performance
in~\cite{ICRC2019-ctaperformance}.
Moreover, CTA will provide the capability of surveying large regions of the sky 
thanks to its increased field of view and a superior angular resolution. 
In particular, a deep Galactic plane survey is envisioned to be one of the main deliverables of CTA.

\section{CTA organization}

%Here I will describe the observatory structure
\begin{figure}
\centering
\includegraphics[width=0.8\textwidth]{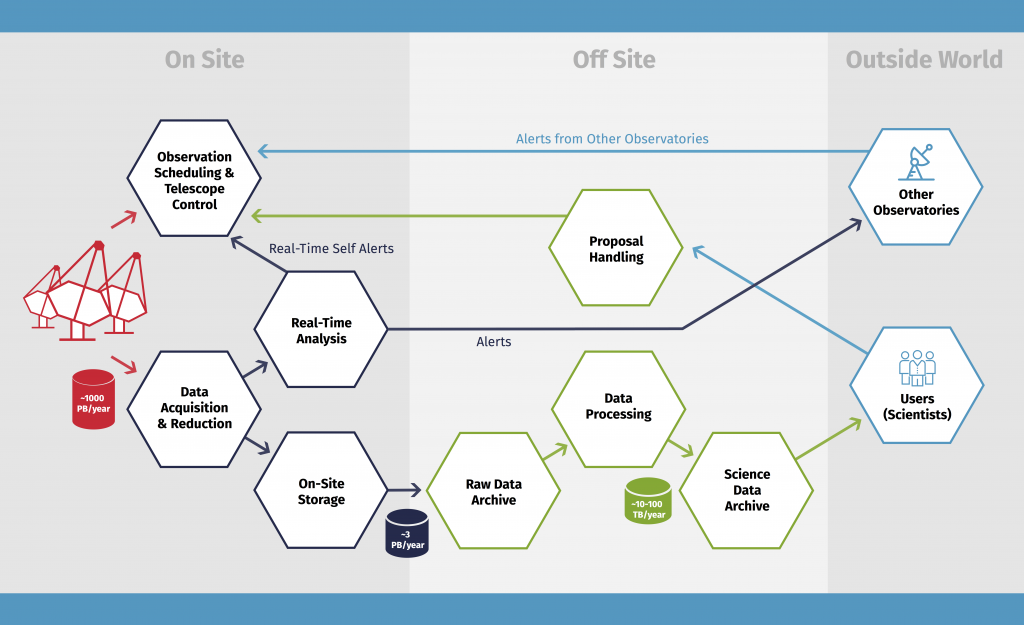}     % includes figure foo.eps
\caption{Structure of the CTA observatory, its
data flow, and interaction with users and other observatories.}
\label{fig:dataflow}
\end{figure}

The CTA observatory 
will be open to a wide community of scientific users from
astronomy and astrophysics, astroparticle physics, particle physics, cosmology
and plasma physics. The data flow and interactions with the users is sketched in Figure~\ref{fig:dataflow}.
Several modes of user access to observation time and to
data products will be provided by the CTA Observatory:
\begin{itemize} 
\item The Guest Observer (GO) Programme by which users can obtain access to
proprietary observation time.
%, submitting proposals in response to Announcements
%of Opportunity (AOs). 
Typical requests per proposal will require 2 -- 100\,h of observation time.
\item The Key Science Projects (KSPs) are large programmes that ensure that the
key science issues for CTA are addressed in a coherent fashion, and generate
legacy data products. KSPs typically require 100 -- 1000\,h of
observation time.
\item Director's Discretionary Time (DDT) represents a small fraction of
observation time reserved for, for example, unanticipated targets of
opportunity.
\item Archive Access under which all CTA gamma-ray data will be openly
available, after a proprietary period.
\end{itemize}

To allow construction and operation of such large scientific infrastructure CTA combines: 
\begin{itemize} 
 \item The CTA observatory (CTAO), which is a legal entity to manage the construction and operation of the observatory.
CTAO is currently a German gGmbH,
and will be transformed into a European Research Infrastructure Consortium (ERIC)
under the European Law, open to countries to be full members from inside and outside of the European Union.
 \item 
The CTA Consortium (CTAC), formed by the scientists and engineers that initiated and designed the instruments for the observatory, 
and who will deliver to CTAO most of the hardware and software that is needed to implement CTA. 
Consortium members are responsible for preparing the science case,
the design, construction and commissioning of the CTA telescopes,
optimizing the array layout, and development of the data analysis chain and calibration scheme. 
As of June 2019, CTAC consists of about 1500 members from 31 countries and 206 institutes.
\end{itemize}

\section{Key Science Projects}

%Here I will describe the Key Science Projects
\begin{figure}
\centering
\includegraphics[width=0.9\textwidth]{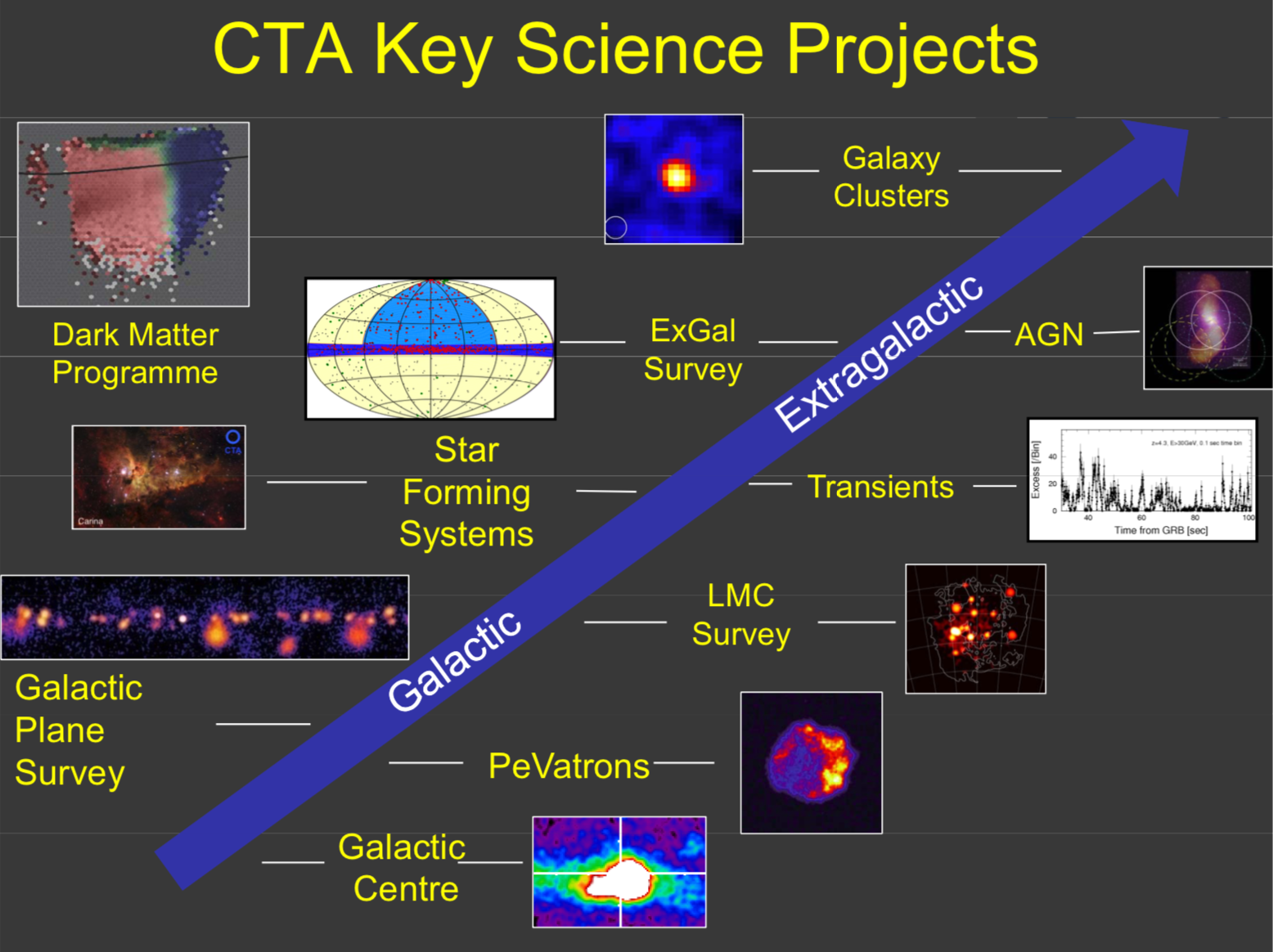}     % includes figure foo.eps
\caption{Key science topics of CTA.}
\label{fig:science}
\end{figure}

The Science with CTA is presented in detail in \cite{CTA-Science:2019}. Summarizing, 
CTA will address major questions in and beyond astrophysics which can be classified under three broad themes:
\begin{itemize}
 \item {\bf Origin and Role of Relativistic Cosmic Particles.}
What are the sites of high-energy particle acceleration in the universe?
What are the mechanisms for cosmic particle acceleration?
What role do accelerated particles play in feedback for star formation and galaxy evolution?
 \item {\bf Probing Extreme Environments.}
What physical processes are at work close to neutron stars and black holes?
What are the characteristics of relativistic jets, winds and explosions?
 \item {\bf Exploring Frontiers in Fundamental Physics.} 
What is the nature of Dark Matter? How is it distributed?
Are there quantum gravitational effects on photon propagation?
Do axion-like particles exist? 
\end{itemize}
Proposed CTA Key Science topics are sketched in Figure~\ref{fig:science} sorted in distance to objects from Earth.
Within the Key Science Projects (KSPs), deep survey fields will be obtained for some key regions
hosting prominent targets, while wider field surveys will be conducted to build
up unbiased population samples and to search for the unexpected. The
combination of the wide CTA FoV with unprecedented sensitivity ensures that CTA
can deliver surveys 1-2 orders of magnitude deeper
than existing surveys within a relatively short amount of observing time 
(few years once the CTA construction is complete).
The CTA surveys will open up
discovery space in an unbiased way and generate legacy datasets of long-lasting
value. 
Several science projects of the CTA consortium can be found in this conference,
%see \cite{ICRC2019-transient,ICRC2019-galacticcenter,ICRC2019-neutrino,ICRC2019-pevatron,ICRC2019-MWLMM,ICRC2019-cosmology}.
see \cite{ICRC2019-transient,ICRC2019-galacticcenter,ICRC2019-MWLMM,ICRC2019-cosmology}.

\section{Status of the telescope prototypes}
%Here I will describe the status of the project
\begin{figure}
\centering
\includegraphics[width=1.0\textwidth]{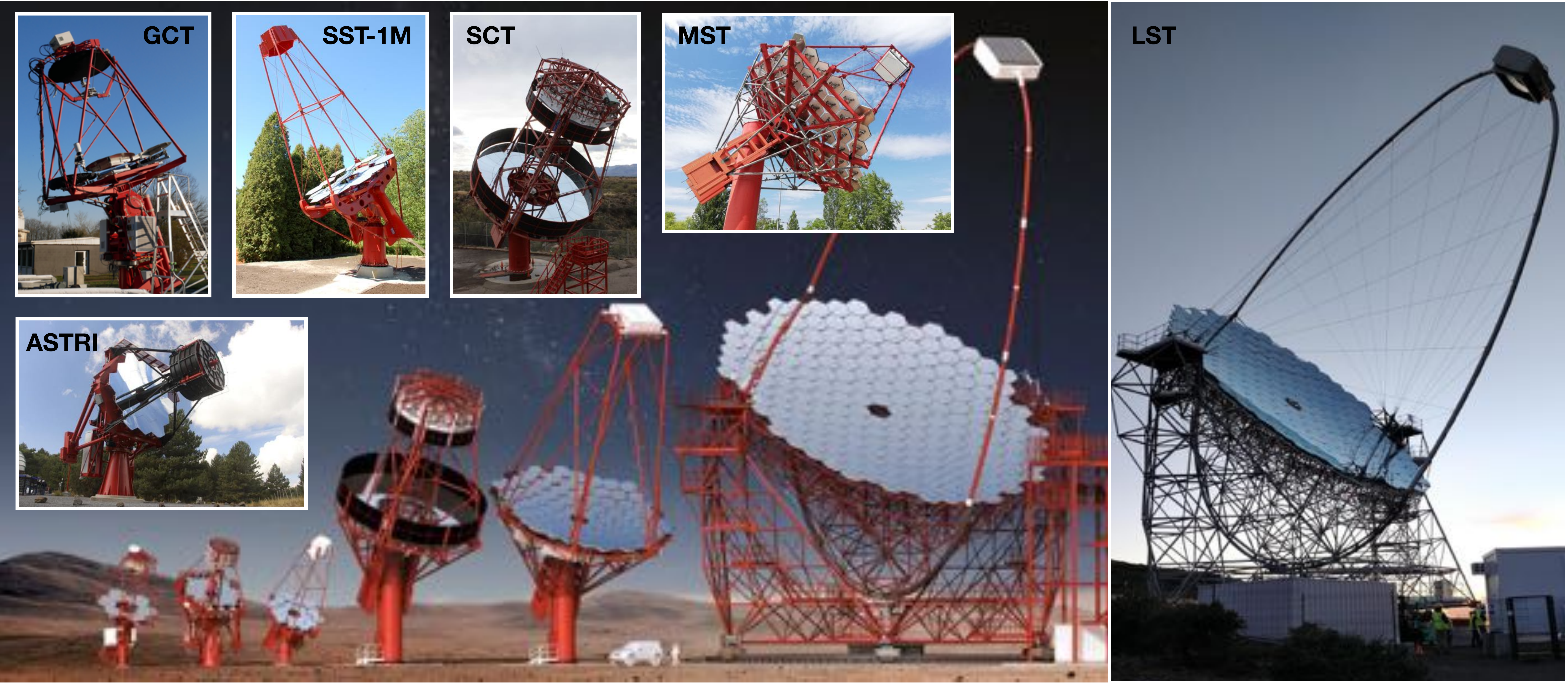}     % includes figure foo.eps
\caption{
 Prototypes for the three sizes of the CTA telescopes:
Large-Sized Telescope (LST), Medium-Sized Telescope (MST), Schwarzschild-Couder Telescope (SCT, which is an alternative design for the MST),
and three Small-Sized Telescopes (SSTs): single mirror SST-1M as well as two dual-mirror designs of SST-2M GCT and SST-2M ASTRI. 
An artists view of the telescopes is in the background while pictures of the actually built prototypes are in the foreground.}
%(a) The LST prototype which is under construction 
%in La Palma. 
%(b) MST prototype in Berlin. 
%(c) SST-1M prototype in Cracow, 
%(d) SST-2M GCT prototype in Paris, 
%and (e) SST-2M ASTRI prototype in Sicily. 
\label{fig:telescopes} 
\end{figure}

In a decade-long effort, CTA Consortium members have developed and built prototype telescopes, and validated
their performance.
Figure~\ref{fig:telescopes} depicts different prototype designs as well as photographs of actually
constructed prototypes. 

For the SST, three different proposals exist: a single mirror dish (SST-1M, \cite{ICRC2019-SST-1M}), with a prototype built
in Krakow, Poland; a dual-mirror design dubbed GCT, with a prototype in Meudon, France; and a 
dual-mirror design dubbed ASTRI \cite{ICRC2019-ASTRI}, with a prototype in Sicily, Italy. Compact SST cameras will be equipped
with photosensors based on Silicon photomultipliers (see e.g. \cite{ICRC2019-CHEC-S}). Currently, CTA is undergoing a harmonization process
with the goal to converge on a single SST design for the production phase.

For the MST, a prototype structure was built and erected in Adlershof, Berlin, Germany.
Two different camera designs, NectarCAM and FlashCam, both based on traditional
photomultiplier tubes, are developed for this telescope type
and both were successfully built, installed and tested on the MST prototype structure in Berlin~\cite{ICRC2019-Nectarcam}.
A first ever dual-mirror Schwarzschild-Couder Telescope
as an alternative design for medium gamma-ray energies, with a compact camera based on Silicon photomultipliers, 
was developed and built in the USA~\cite{ICRC2019-SCT},
and inaugurated in Arizona at a basecamp of the Whipple observatory in January 2019.

For the LST, a prototype telescope LST1 was constructed at the final CTA site at the observatory Roque de los Muchachos
in La Palma, Spain \cite{ICRC2019-LST}. LST1 was inaugurated in October 2018 and is now undergoing a commissioning phase
\cite{ICRC2019-LSTCAM}
with a target to become the first accepted and operational CTA telescope by the end of 2020.

\section{Status of CTA observatory}

CTAO has the headquarters in Bologna, Italy and is ramping up its staff members. 
The CTA Science Data Management Centre (SDMC), 
which will make CTA science data products available to the worldwide community, 
will be located in a new building complex on the Deutsches
Elektronen-Synchrotron (DESY) campus in Zeuthen, just outside Berlin. % (see the design of the building in Figure~\ref{fig:sdmc}).
A competition to design and construct a new building was initiated by DESY in
2018. The final decision on the winning design, with an award for the first,
second and third winner, was announced in March 2019. 

\begin{figure}
\centering
\includegraphics[width=0.52\textwidth]{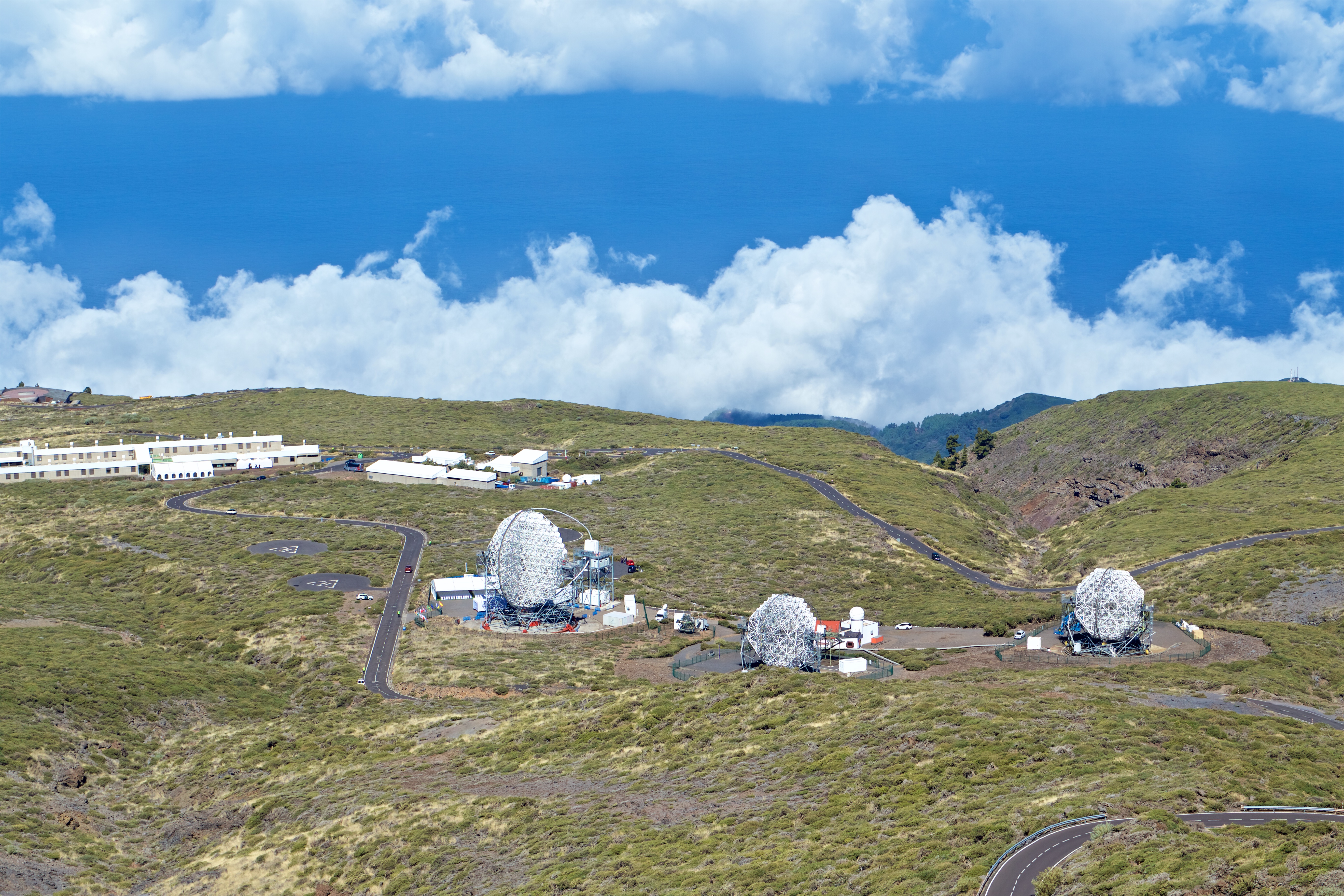}     % includes figure foo.eps
\includegraphics[width=0.47\textwidth]{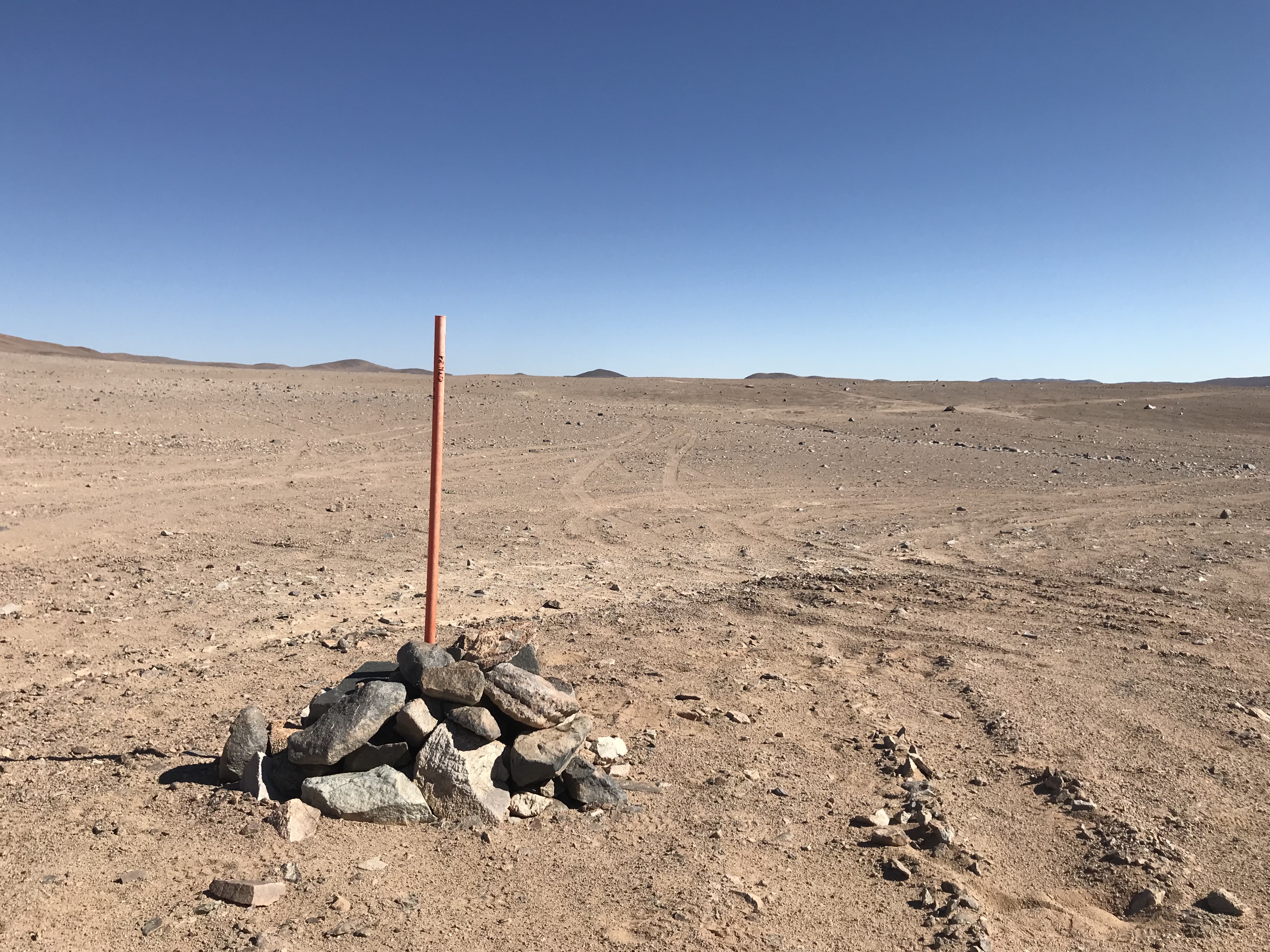}     % includes figure foo.eps
\caption{Pictures of the CTA sites, CTA-North in La Palma (left) and CTA-South in Paranal (right).}
\label{fig:sites}
\end{figure}

The CTA site selection process was successfully finished and hosting agreements were
signed with the European Southern Observatory (ESO) and Chilean government for the Paranal site in Chile and
with the Instituto de Astrofisica de Canarias (IAC) for the Roque de los Muchachos
Observatory in La Palma, Spain. 
The CTA-South array will be operated by ESO on behalf of CTAO.
Infrastructure implementation planning is ongoing and it is foreseen to start deployment in Chile in 2020.
For CTA-North, while the LST1 prototype is under commissioning,
preparations are ongoing for the next stage infrastructure to cover
LST2, 3 and 4 as well as one MST and an operation building. The current plan foresees to build 
this infrastructure and the telescopes by 2022.
After a technical commissioning of these telescopes, early science
can be expected with partial arrays as early as in 2022 while the entire
arrays should be completed by 2025.  % and unfold their science potential by that time.  
The pictures of the CTA sites as of beginning of 2019 can be seen in Figure~\ref{fig:sites}.

Operations of CTA are foreseen to last for 30
years.  CTA perspectives in relativistic astrophysics include the in-depth
understanding of known VHE gamma-ray emitters and their mechanisms,
detection of new object classes, and discovery of new phenomena -- as for
many facilities breaking new ground, the most important discoveries may not
be the ones discussed in today's science case documents.

\acknowledgments
CTA gratefully acknowledges support from the agencies and organizations listed under Funding Agencies at this website: 
http://www.cta-observatory.org/.

%\vspace{-0.3cm}

%CTA Science Case Volume of the CTA Technical Design Report, J. Hinton (Ed.), OBS-TDR/141106 (2015),
%to be published

%%


\begin{thebibliography}{99}
\bibitem{weekesCrab} 
Weekes T.C., Cowley M.F., Fegan D.J. et al., 
\emph{Observation of TeV Gamma Rays from the Crab Nebula Using the Atmospheric Cerenkov Imaging Technique},
\emph{ApJ} {\bf 342} 379ff (1989).
\bibitem{deNauroisMazin:2015} 
{de Naurois M. \& Mazin D.},
\emph{Ground-based detectors in very-high-energy gamma-ray astronomy},
{Comptes rendus - Physique} {\bf 16} 610ff (2015).
\bibitem{Degrange:2015} 
{Degrange B. \& Fontaine G.},
\emph{Introduction to high-energy gamma-ray astronomy},
{Comptes rendus - Physique} {\bf 16} 587ff (2015). 
\bibitem{ICRC2019-ctaperformance} 
{Maier, G. et al.},
\emph{Performance of the Cherenkov Telescope Array},
\emph{Proceedings of 36th ICRC} 2019, this volume. 
\bibitem{ICRC2019-transient} 
{Sch\"ussler, F. et al.},
\emph{The Transient program of the Cherenkov Telescope Array},
\emph{Proceedings of 36th ICRC} 2019, this volume. 
\bibitem{ICRC2019-galacticcenter} 
{Viana, A. et al.},
\emph{The Cherenkov Telescope Array view of the Galactic Center region},
\emph{Proceedings of 36th ICRC} 2019, this volume. 
%\bibitem{ICRC2019-neutrino} 
%{Satalecka, K. et al.},
%\emph{Proceedings of 36th ICRC} 2019, this volume. 
%\bibitem{ICRC2019-pevatron} 
%{Ang\"uner, E.O. et al.},
%\emph{Proceedings of 36th ICRC} 2019, this volume. 
\bibitem{ICRC2019-MWLMM} 
{Barres de Almeida, U. et al.},
\emph{Cherenkov Telescope Array Science: A multi-wavelength and multimessenger perspective},
\emph{Proceedings of 36th ICRC} 2019, this volume. 
\bibitem{ICRC2019-cosmology} 
{Martinez-Huerta, H. et al.},
\emph{Testing cosmology and fundamental physics with the Cherenkov Telescope Array},
\emph{Proceedings of 36th ICRC} 2019, this volume. 
\bibitem{ICRC2019-SST-1M} 
{Heller, M. et al.},
\emph{The SST-1M project for Cherenkov Telescope Array},
\emph{Proceedings of 36th ICRC} 2019, this volume. 
%\bibitem{ICRC2019-GCT} 
%{Dmytriiev, A. et al.},
%\emph{Proceedings of 36th ICRC} 2019, this volume. 
\bibitem{ICRC2019-ASTRI} 
{Pareschi, G. et al.},
\emph{The ASTRI program in the context of the Cherenkov Telescope Array Observatory},
\emph{Proceedings of 36th ICRC} 2019, this volume. 
\bibitem{ICRC2019-CHEC-S} 
{Watson, J. et al.},
\emph{Commissioning and Performance of CHEC-S -- a compact high-energy camera for the Cherenkov Telescope Array},
\emph{Proceedings of 36th ICRC} 2019, this volume. 
%\bibitem{ICRC2019-MST} 
%{Barbosa Martins, V. et al.},
%\emph{Proceedings of 36th ICRC} 2019, this volume. 
\bibitem{ICRC2019-Nectarcam} 
{Glicenstein, J.-F. et al.},
\emph{Status of the Davies Cotton and Schwarzschild-Coude Medium-Sized Telescopes for the Cherenkov Telescope Array},
\emph{Proceedings of 36th ICRC} 2019, this volume. 
\bibitem{ICRC2019-SCT} 
{Vassiliev, V. et al.},
{\textit {Prototype 9.7m Schwarzschild-Couder telescope for the Cherenkov Telescope Array,}}
\emph{Proceedings of 36th ICRC} 2019, this volume. 
\bibitem{ICRC2019-LST} 
{Cortina, J. et al.},
\emph{Status of the Large Size Telescopes of the Cherenkov Telescope Array},
\emph{Proceedings of 36th ICRC} 2019, this volume. 
\bibitem{ICRC2019-LSTCAM} 
{Sakurai, S. et al.},
{\textit {The calibrations of the first Large Sized Telescope of the Cherenkov Telescope Array,}}
\emph{Proceedings of 36th ICRC} 2019, this volume. 
%\bibitem{CTAspecialIssue:2013} 
%{Hinton J., Sarkar S., Torres D. \& Knapp J. (Eds.)},
%{AP} 2013 {\bf(43)} 1ff.   
\bibitem{CTA-Science:2019} 
CTA Consortium: Acharya B.S. et al., 
\emph{Science with the Cherenkov Telescope Arra}, World Scientific Publishing,  doi:10.1142/10986 (2019)
%doi:10.1142/10986, arXiv:1709.07997
\end{thebibliography}
\end{document}